\documentclass[prl,twocolumn,superscriptaddress,showpacs,floatfix]{revtex4}
\usepackage{epsfig}
\usepackage{amsfonts, amssymb}
\newcommand{\tc}{,~}

\begin{document}

\title{\boldmath A Pair Polarimeter for Linearly Polarized High Energy Photons} 

\author{C.~de~Jager}
\affiliation{\mbox{Thomas Jefferson National Accelerator Facility\tc Newport News, VA 23606}}

\author{B.~Wojtsekhowski}
\affiliation{\mbox{Thomas Jefferson National Accelerator Facility\tc Newport News, VA 23606}}

\author{D.~Tedeschi}
\affiliation{\mbox{University of South Carolina, Columbia, SC 29208, USA}}

\author{B.~Vlahovic}
\affiliation{\mbox{Thomas Jefferson National Accelerator Facility\tc Newport News, VA 23606}}
\affiliation{\mbox{North Carolina Central University, Durham, NC 27707, USA}}

\author{D.~Abbott}
\affiliation{\mbox{Thomas Jefferson National Accelerator Facility\tc Newport News, VA 23606}}

\author{J.~Asai}
\affiliation{\mbox{Canadian Light Source, University of Saskatchewan, Saskatoon, Canada}}

\author{G.~Feldman}
\affiliation{\mbox{The George Washington University, Washington, DC 20064, USA}}

\author{T.~Hotta}
\affiliation{\mbox{Research Center for Nuclear Physics, Osaka University, Ibaraki, Osaka 567-0047, Japan}}

\author{M.~Khadaker}
\affiliation{\mbox{Norfolk State University, Norfolk, VA 23504, USA}}

\author{H.~Kohri}
\affiliation{\mbox{Research Center for Nuclear Physics, Osaka University, Ibaraki, Osaka 567-0047, Japan}}

\author{T.~Matsumara}
\affiliation{\mbox{Research Center for Nuclear Physics, Osaka University, Ibaraki, Osaka 567-0047, Japan}}

\author{T.~Mibe}
\affiliation{\mbox{Research Center for Nuclear Physics, Osaka University, Ibaraki, Osaka 567-0047, Japan}}

\author{T.~Nakano}
\affiliation{\mbox{Research Center for Nuclear Physics, Osaka University, Ibaraki, Osaka 567-0047, Japan}}

\author{V.~Nelyubin}
\affiliation{\mbox{Thomas Jefferson National Accelerator Facility\tc Newport News, VA 23606}}
\affiliation{\mbox{University of South Carolina, Columbia, SC 29208, USA}}

\author{G.~Orielly}
\affiliation{\mbox{The George Washington University, Washington, DC 20064, USA}}

\author{A.~Rudge}
\affiliation{\mbox{CERN EP Division, 121 Geneva 23, Switzerland}}

\author{P.~Weilhammer}
\affiliation{\mbox{CERN EP Division, 121 Geneva 23, Switzerland}}

\author{M.~Wood}
\affiliation{\mbox{University of South Carolina, Columbia, SC 29208, USA}}

\author{T.~Yorita}
\affiliation{\mbox{Research Center for Nuclear Physics, Osaka University, Ibaraki, Osaka 567-0047, Japan}}

\author{R.~Zegers}
\affiliation{\mbox{Japan Atomic Energy Research Institute, Mikazuki, Hyogo 679-5148, Japan}}

\date{Received: 1 August / Accepted: 14 Nov 2003}

\begin{abstract}
{A high quality beam of linearly polarized photons of several GeV will become
available with the coherent bremsstrahlung technique at JLab. 
We have developed a polarimeter which requires about two meters of the beam line,
has an analyzing power of 20\% and an efficiency of 0.02\%.
The layout and first results of a polarimeter test on the laser back-scattering 
photon beam at SPring-8/LEPS  are presented.}
\end{abstract}


\maketitle


\section{Introduction}
\label{sec:introduction}
Electron-positron pair photo-production from amorphous matter can be calculated exactly 
in the framework of QED \cite{Max59}. 
The azimuthal distribution in this reaction can be presented as
\mbox{$\sigma = \sigma_{unpol}\left[1 \,+\, P_l \cdot A \cdot cos(2\phi)\right]$}, 
where $P_l$ is degree of photon linear polarization, $A$ the analyzing power 
of the process and $\phi$ the azimuthal angle between the polarization plane and the detected
particle(s). The value of $A$ ranges from 0.1 to 0.3 depending on the reaction kinematics. 
The use of the $\gamma,e^+e^-$ reaction in polarimetry was suggested in 1950 \cite{You50}. 
The main experimental challenge in such a polarimeter is the small angle between the pair 
components. This was resolved by using a magnetic field to separate the 
electron and positron \cite{Bar62} at the price of a reduced analyzing power.
For photons in the few GeV energy range a different solution was 
proposed \cite{Woj98} based on a silicon microstrip detector, 
of which the unique high track resolution allows the observation
a $e^+e^-$ pair with a separation as small as \mbox{100 $\mu$m.}

\section{Kinematics}
\label{sec:kinematics}

\begin{figure}[htp]
\begin{center}
	\epsfig{figure=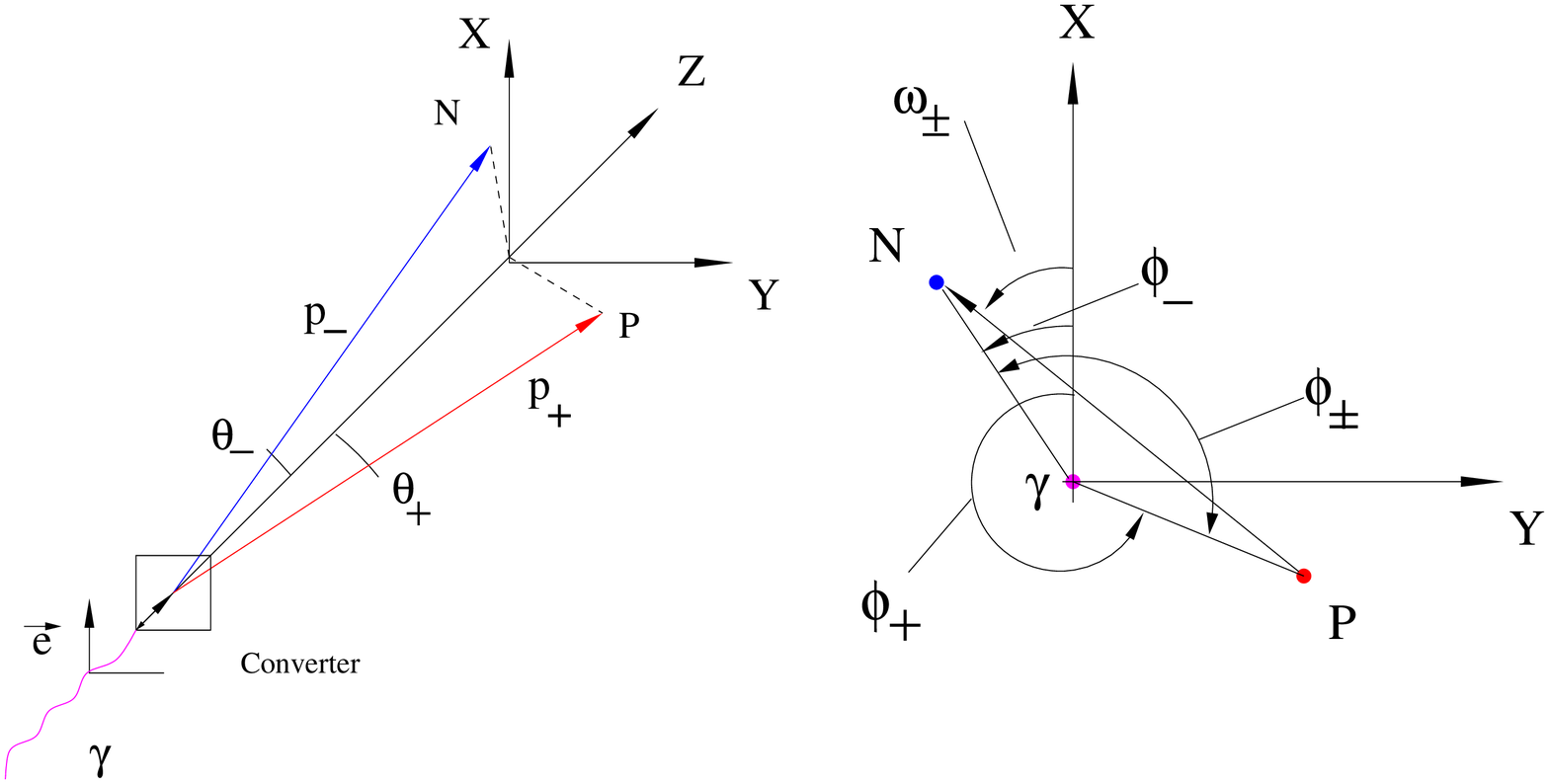,width=8.5cm}
\end{center}
	\caption{The kinematics of $e^+e^-$ pair photo production (left picture) 
and the azimuthal angles in the detector plane. The photon momentum is directed along the Z axis. 
The photon polarization vector $\vec e$ is parallel to the X axis. 
The angle $\phi_+$($\phi_-$) is the angle between the photon polarization 
plane and the plane constructed by the momentum of the photon and the 
momentum of the positron (the electron). 
The angle $\phi_{\pm}$ is called the coplanarity angle. 
The labels P and N indicate the positions of the crossings of the detector
plane by the positron and the electron. 
The angle $\omega_{\pm}$ between the polarization plane and the vector $\overline{PN}$ 
is a directly measurable parameter.} 
	\label{fig:kinem}
\end{figure}

Figure \ref{fig:kinem} illustrates the kinematics of electron-positron pair production 
and its relevant variables. The use of the angle $\omega_\pm$, which is the azimuthal angle of 
the $e^+e^-$ plane, yields a value of $A \sim 0.24$ for the case of a very thin converter 
and equal energies of the electron and positron. 

In the determination of the asymmetry we used a vector $\bf n_\omega $ constructed
from the momenta of the electron($\bf p_+$), the positron($\bf p_-$), and the photon($\bf k$) as

${\bf n}_\omega = \frac { {\bf p_{+\perp}/|p_{+\parallel}|} - {\bf p_{-\perp}/|p_{-\parallel}|} } 
                       {|{\bf p_{+\perp}/|p_{+\parallel}|} - {\bf p_{-\perp}/|p_{-\parallel}|}|}$ 

where ${\bf p_{\pm\perp} = p_\pm - p_{\pm\parallel} }$, ${\bf p_{\pm\parallel} = p_\pm \cdot k} /k$.
This vector lies in the plane perpendicular to the photon momentum {\bf k}. 
The direction of the {\bf n}$_\omega$ vector is determind from 
the positron and the electron tracks in the detector, 
even the position of the pair creation vertex is not detected.

From the azimuthal dependence of asymmetry we find $A_{exp}$:
\begin{equation}
\frac{N(\omega,\parallel) \,-\, N(\omega+\frac{\pi}{2},\perp)}{N(\omega,\parallel) 
\,+\, N(\omega+\frac{\pi}{2},\perp)}\,=\, A_{exp}\, cos(2\omega + \Delta),
	\label{eq:effect}
\end{equation}
where \mbox{$N(\omega,\parallel)$} is the number of $e^+e^-$ pairs observed at the 
angle $\omega$ for the photon polarization {\bf e} parallel to the vertical and 
\mbox{$N(\omega+\frac{\pi}{2},\perp)$} is the number of pairs observed at the 
angle $\omega+\frac{\pi}{2}$ for the horizontal photon polarization, 
$\Delta$ is the tilt of the detector axis.


\section{\boldmath Experiment}
\label{sec:setup}

\subsection{Layout}
\label{sec:layout}

\begin{figure}[htp]
\begin{center}
        \epsfig{figure=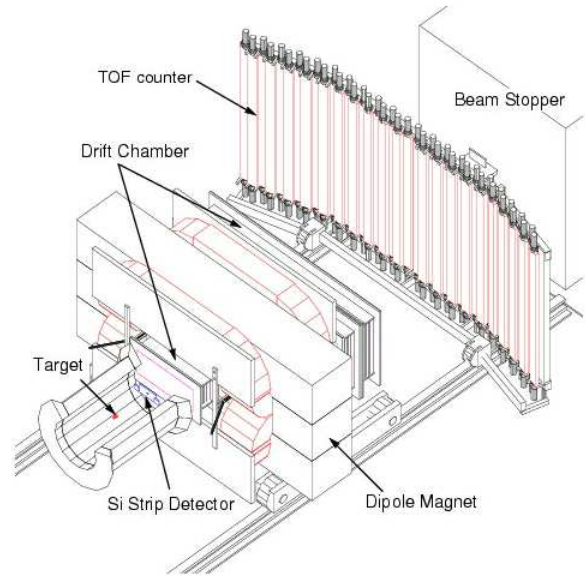,angle=0.,width=8.5cm}
\end{center}
        \caption{The LEPS spectrometer.}
        \label{fig:leps}
\end{figure}

The experiment was performed at the SPring-8 facility in Japan. SPring-8 is a third-generation synchrotron 
radiation facility with 62 beam lines for various applications. 
One of the beam lines with high-energy photons is used for high-energy nuclear physics.
The photons are produced by backward Compton scattering laser light off 
the electrons circulating in the storage ring. 
The beam line is called LEPS (Laser Electron Photon beam line at SPring-8) \cite{Nak01}.
The laser system focuses the light at the intersection region located between 
two bending magnets in the storage ring. An Argon laser provides the laser beam with a wave 
length of 351 nm for which the maximum energy of the produced photons is 2.4 GeV. 
The tagging system employs the bending magnet located down stream of the interaction region
for analyzing the momentum of the electrons. 
The energy resolution of the tagging system is about 15 MeV in the region from 1.5 to 2.5 GeV.
The produced photons are led to the experimental hutch located \mbox{69 m} away from the 
interaction region. The beam spot size on the target is about \mbox{2 cm}. 
Figure~\ref{fig:leps} shows the LEPS spectrometer in the experimental hutch. The main 
components of LEPS are a dipole magnet, a set of drift chambers and a time-of-flight wall.    

\begin{figure}[htp]
\begin{center}
        \epsfig{figure=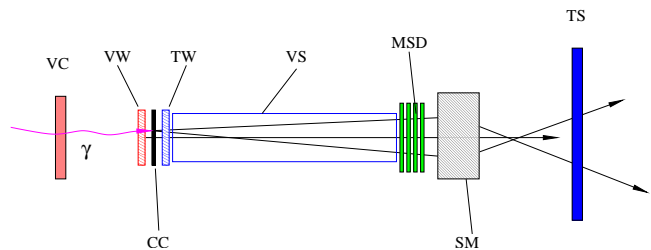,angle=0.,width=8.5cm}
\end{center}
        \caption{The layout of the photon polarimeter. 
The photon arrives from the left. 
The veto detector is marked as VC, the veto wire chamber as VW, 
the converter as CC, the trigger wire chamber as TW, the vacuum straight section as VS, 
the set of micro-strip detectors as MSD, the separation magnet as SM, 
and the trigger scintillator counter as TS.}
        \label{fig:layout}
\end{figure}

The polarimeter scheme is presented in Fig.~\ref{fig:layout}. 

The polarimeter was installed in front of the LEPS spectrometer at the position
used by the target. The LEPS drift chambers were used for detecting the electron and 
the positron. The magnet SM together with the  LEPS drift chambers formed a pair spectrometer.
The LEPS dipole magnet was turned off.

\subsection{Photon beam}
\label{sec:photons}

\begin{figure}[htp]
        \epsfig{figure=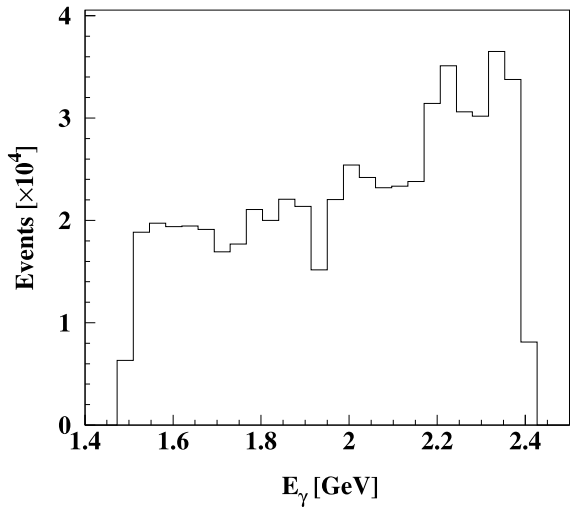,angle=0.,width=7.5cm}
        \caption{The tagged photon energy spectrum.}
        \label{fig:pict5}
\end{figure}

\begin{figure}[htp]
        \epsfig{figure=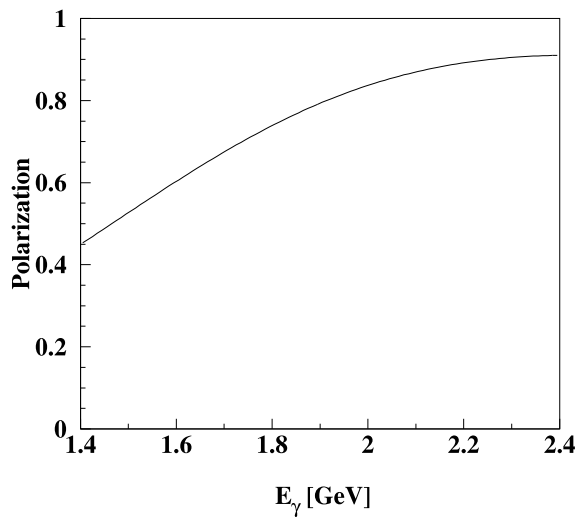,angle=0.,width=7.5cm}
        \caption{The photon polarization as a function of the photon energy.}
        \label{fig:polar}
\end{figure}

Figure~\ref{fig:pict5} shows the photon energy spectrum detected by the tagger. 
The calculated photon polarization vs photon energy is shown in Fig.~\ref{fig:polar}. 
It has a maximum value of 93\% at the Compton edge of 2.4 GeV. 
The energy dependence of the polarization was taken into account in the extraction 
of the analyzing power (see section~\ref{sec:results}).

\subsection{Trigger logic}
\label{sec:trigger}

The signals from VC, WV, TW and TS are used to form a polarimeter trigger signal (PT).
The coincidence between PT and a tagger trigger signal was used to initiate 
a readout cycle of both the polarimeter and the LEPS electronics.
The amplitude from the trigger wire chamber TW was recorded in both DAQs 
and used to cross check the synchronization of information.  

\subsection{Data collection}
\label{sec:data}
Measurements were done at two orientations of the laser polarizations - vertical and horizontal.
The orientations were interchanged typically every 2 hours. 
The degree of the laser photon polarization was 99\%. 
The typical tagger rate was 150 kHz, the wire chamber rates 20 kHz and the
event rate 150 Hz. In the trigger rate the contribution from the unpolarized 
background of photons produced on the residual gas in the storage ring was below 2\%.   

\section{\boldmath Analysis and Results}
\label{sec:analysis_results}

\subsection{Data analysis}
\label{sec:analysis}

Only  events with one hit in the tagger were considered.
For reconstruction of the tracks in the MSD and the drift chambers first 
clusters were searched. 
After a cluster was recognized and its position determined, the hit pattern was 
used for a track search. 
We analyzed all combinations and selected two with best matching of the hit positions. 
The distribution of the residuals for the best combination has a peak, of which width allows 
to estimate the MSD spatial resolution to be 15~$\mu$m. 
Events with two tracks in the MSD were used in the analysis of the pair angular and position 
distributions. The drift chamber and tagger information were used to calibrate the
deflection parameter of the pair spectrometer. 
The energy of the electron and the positron were used to select pairs with $E_+ \sim E_-$. 
The separation between the electron and the positron tracks in the MSD ($\Delta r$) and the photon 
energy $E_\gamma$ were used to calculate the reduced open angle 
$\Theta = \Delta r/L \cdot E_\gamma/m_e$, where $L$ is the distance between
the converter and MSD, $m_e$ the electron mass. 

\subsection{Results}
\label{sec:results}

\begin{figure}[htp]
\begin{center}
        \epsfig{figure=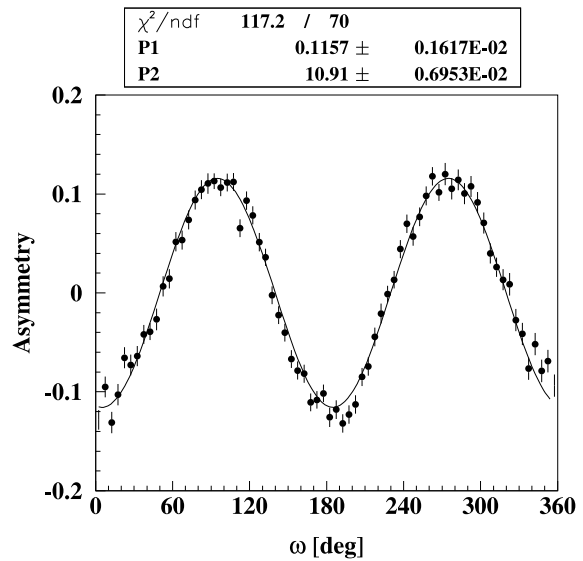,angle=0.,width=7.5cm}
\end{center}
        \caption{
The beam polarization effect in the azimuthal distribution of $e^+e^-$ pairs. 
$\omega$ is the angle between the $e^+e^-$ plane and the photon polarization plane. 
Parameters P1 and P2 of the fit reflect the asymmetry $A_{exp}$ and $\Delta$ in 
eq.~\ref{eq:effect}.}
        \label{fig:pict3}
\end{figure}

The beam polarization effect in the azimuthal distribution (see eq.~\ref{eq:effect})
of the pair plane is shown in Fig.~\ref{fig:pict3} for events with a photon energy 
in the range 1.5-2.4 GeV.

\begin{figure}[htp]
\begin{center}
        \epsfig{figure=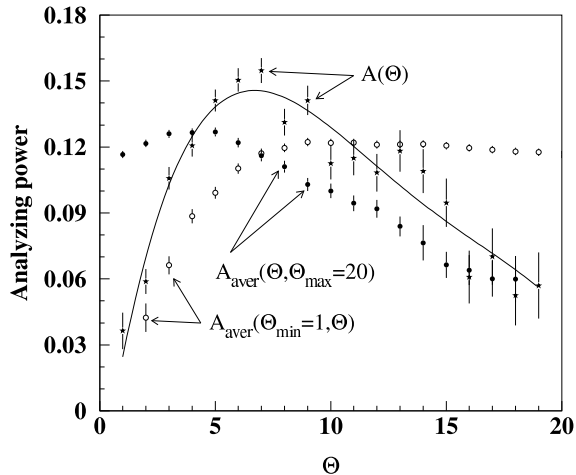,angle=0.,width=7.5cm}
\end{center}
        \caption{The analyzing power $A$ as a function of the reduced open angle $\Theta$. 
The solid line shows the trend of the experimental data (stars). 
The average value (open circles) $A_{aver}(\Theta_{min}=1,\Theta)$
shows the effect of a running upper limit for a fixed value of a low limit.
The average value (closed circles) $A_{aver}(\Theta_{min}=\Theta, 20)$
shows the effect of a running lower limit for a fixed value of upper limit.} 
        \label{fig:pict4}
\end{figure}

The analyzing power of the reaction has a strong dependence on the open angle between 
the pair components.
Figure~\ref{fig:pict4} shows a corresponding experimental result. 
Two cases of the averaged analyzing power are also presented in this plot. 
These running averagies show how to select the low and the upper limits on 
the reduced open angle to minimize the sensitivity of the analyzing power $A$ to these cuts.

\begin{figure}[htp]
\begin{center}
        \epsfig{figure=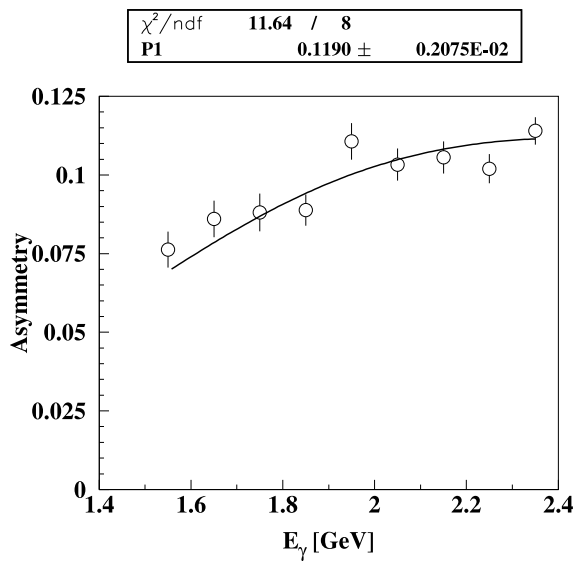,angle=0.,width=7.5cm}
\end{center}
        \caption{The asymmetry $A_{exp}$ as a function of the photon energy for all pairs.
The shape of the curve follows the polarization of the photon beam.}
        \label{fig:pict2}
\end{figure}

For the next step in the analysis we used only events with a reduced open angle 
in the range from 4 to 20.
The asymmetry as a function of the photon energy is plotted in Fig.~\ref{fig:pict2}.
The curve follows the dependence of the photon beam polarization as a function 
of the photon energy.

\begin{figure}[htp]
\begin{center}
        \epsfig{figure=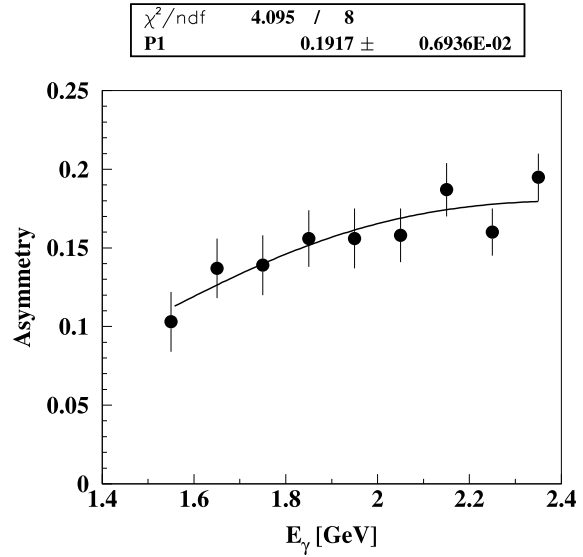,angle=0.,width=7.5cm}
\end{center}
        \caption{The asymmetry $A_{exp}$ as a function of the photon energy for the pairs
with a condition $\,0.8 <\, E_+ / E_- \,< 1.2 $.}
        \label{fig:pict1}
\end{figure}

The selection of the pairs with almost equal $e^+$ and $e^-$ energies leads to 
a higher asymmetry as is shown in Fig.~\ref{fig:pict1}. 
The values of the analyzing power for these two cases are 0.119 and 0.192. They agree 
within 5\% with calculations \cite{Woj03} for the thickness of the carbon converter
used in our experiment (0.1mm). 

\section{Conclusion}
\label{sec:conclusion}

Here, we described the results of a polarimeter test at SPring-8/LEPS. 
For the first time the polarization effect in the pair photo-production 
from amorphous matter was observed in the GeV energy range. 
We demonstrated that a compact polarimeter for photon energies of several GeV 
can be constructed by using silicon micro-strip detectors.
An analyzing power was observed of 0.192
for 0.1 mm carbon converter and near equal energies of 
the electron and the positron, which is within 5\% of the expected value.

We acknowledge the crucial support of W.~Briscoe, L.~Cardman, and B.~Mecking. 

This work was supported in part by the National Science Foundation in grants 
PHY-0099487 for the North Carolina Central University and PHY-0072361 
for the University of South Carolina and by DoE contract DE-AC05-84ER40150
under which the Southeastern Universities Research Association (SURA) operates
the Thomas Jefferson National Accelerator Facility for the United States
Department of Energy.

\end{document}